\documentclass[singlecolumn,showpacs,preprintnumbers,amsmath,amssymb,superscriptaddress]{revtex4}
\usepackage{graphicx}
\usepackage{dcolumn}
\usepackage{bm}
\usepackage{longtable}

\newcommand{\bea}{\begin{eqnarray}}
\newcommand{\eea}{\end{eqnarray}}

\usepackage{amsfonts}
\usepackage{amssymb}
\usepackage{amsmath}
\usepackage{amsthm}
\usepackage{graphicx}

\newcommand{\be}{\begin{equation}}
\newcommand{\ee}{\end{equation}}

\def\Be'{\beta_\mu^{'}}

\def\<{\bigl\langle}
\def\>{\bigr\rangle}

\begin{document}


\title{Collective behaviours: from biochemical kinetics to electronic circuits}

\author{Elena Agliari}
\affiliation{Dipartimento di Fisica, Universit\`{a} degli Studi di Parma, viale G. Usberti 7, 43100 Parma, Italy}
\affiliation{INFN, Gruppo Collegato di Parma, viale G. Usberti 7, 43100 Parma, Italy}
\author{Adriano Barra}
\affiliation{Dipartimento di Fisica, Sapienza Universit\`{a} di Roma, Piazzale Aldo Moro 2, 00185, Roma, Italy}
\author{Raffaella Burioni}
\affiliation{Dipartimento di Fisica, Universit\`{a} degli Studi di Parma, viale G. Usberti 7, 43100 Parma, Italy}
\affiliation{INFN, Gruppo Collegato di Parma, viale G. Usberti 7, 43100 Parma, Italy}
\author{Aldo Di Biasio}
\affiliation{Dipartimento di Fisica, Universit\`{a} degli Studi di Parma, viale G. Usberti 7, 43100 Parma, Italy}
\author{Guido Uguzzoni}
\affiliation{Dipartimento di Fisica, Universit\`{a} degli Studi di Parma, viale G. Usberti 7, 43100 Parma, Italy}

\date{\today}

\begin{abstract}
In this work we aim to highlight a close analogy between cooperative behaviors in chemical kinetics and cybernetics; this is realized by using a common language for their description, that is mean-field statistical mechanics. First, we perform a one-to-one mapping between paradigmatic behaviors in chemical kinetics (i.e., non-cooperative, cooperative, ultra-sensitive, anti-cooperative) and in mean-field statistical mechanics (i.e., paramagnetic, high and low temperature ferromagnetic, anti-ferromagnetic).
Interestingly, the statistical mechanics approach allows a unified, broad theory for all scenarios and, in particular, Michaelis-Menten, Hill and Adair equations are consistently recovered. This framework is then tested against experimental biological data with an overall excellent agreement. One step forward, we consistently read the whole mapping from a cybernetic perspective, highlighting deep structural analogies between the above-mentioned kinetics and fundamental bricks in electronics (i.e. operational amplifiers, flashes, flip-flops), so to build a clear bridge linking biochemical kinetics and cybernetics.
\end{abstract}



 \maketitle

\section*{Introduction}
Cooperativity is one of the most important properties of molecular interactions in biological systems and it is often invoked to account for collective features in binding phenomena.

In order to investigate and predict the effects of cooperativity, chemical kinetics proved to be a fundamental tool and, also due to its broadness over several fields of biosciences, a number of cooperativity quantifiers (e.g. Hills number \cite{hill},  Koshland cooperativity test \cite{koshland}, global dissociation quotient \cite{PCK}, weak and strong fine tunings \cite{enzimi}, etc.), apparently independent or distinct, have been introduced.
However, a clear, unified, theoretical scheme where all cooperative behaviors can be framed would be of great importance, especially in biotechnology research \cite{winfree1,winfree2} and for scientists dealing with interdisciplinary applications \cite{berry,paun}.
To this task, statistical mechanics  offers a valuable approach as, from its basic principles,
it aims to figure out collective phenomena, possibly overlooking the details of the interactions to focus on the very key features.
Indeed, a statistical mechanics description of reaction kinetics has already been paved through theoretical models based on linear Ising chains \cite{thompson}, spin lattices with nearest neighbors interactions \cite{chay}, transfer matrix theory \cite{thompson,chay} and structural probabilistic approaches \cite{wyman}.

In this work we expand such statistical mechanics picture toward a mean-field  perspective \cite{aldo} by assuming that the interactions among the system constituents
are not limited by any topological or spatial constraint, but are implicitly taken to be long-ranged, as in a system that  remains spatially homogeneous. This approach is naturally consistent with the rate-equation picture, typical of chemical kinetics investigations and whose validity is restricted to the case of vanishing correlations \cite{ben,vankampen} and requires a sufficiently high spatial dimension or the presence of an effective mixing mechanism (hence, ultimately, long-range interactions). In general, in the mean-field limit, fluctuations naturally decouple from the volume-averaged quantities and can be treated as negligible noise.
%
%
%

By adopting a mean-field approach, we abandon a direct spatial  representation of binding structures and we introduce a renormalization of the effective couplings. The reward lies in a resulting unique model exhibiting a rich phenomenology (e.g. phase transitions), which low-dimensional models typically lack, yet being still feasible for an exact solution. In particular, we obtain an analytical expression for the saturation function which is successfully compared with recent experimental findings, taken from different (biological) contexts to check robustness. Furthermore, from this theory basic chemical kinetics equations (e.g. Michaelis-Menten, Hill and Adair equations) are recovered as special cases.

Further, there is a deep theoretical motivation underlying the development of a mean-field statistical mechanics approach to chemical kinetics: it can be used to code collective behavior of biosystems into a cybernetical framework. In fact, cybernetics, meant as the science dedicated to the understanding of self-organization and emergent communication among the constituents of a system, can be naturally described via (mean-field) statistical mechanics \cite{MPV,hopfieldnature,AGS}. Thus, the latter provides a shared formalism which allows to automatically translate chemical kinetics into cybernetics and vice versa. In this perspective, beyond theoretical interest, at least two concrete benefits may stem from our investigation: first, in the field of biotechnologies, logical gates have already been obtained through biological hardware (see e.g. \cite{winfree1,winfree2}) and for their proper functioning signal amplification turns out to be crucial. In this paper, cooperativity in reaction kinetics is mapped into amplification in electronics, hence offering a solid scaffold for biological amplification theory.
\newline
Then, as statistical mechanics has been successfully applied in the world of computing (for instance in neural networks \cite{amit}, machine learning \cite{bishop} or complex satisfiability \cite{mezardmontanari}), its presence in the theory of biological processors could be of relevant interest. In particular, we discuss how to map ultra-sensitive kinetics to logical switches and how to read anti-cooperative kinetics as the basic ingredient for memory storage in biological flip-flops, whose interest resides in several biological machineries as gene regulatory networks \cite{enzo}, riboswitches \cite{mandal}, synaptic switches \cite{rovira},  autopoietic systems \cite{autopoiesi} and more \cite{Li-Nature2012,Bonnet-Science2013,Chang-PNAS2012}.

To summarize, a rigorous, promising link between cybernetics and collective biological systems can be established via statistical mechanics and this point will be sketched and corroborated by means of several examples throughout this paper, which is structured as follows:
\newline
First, we review the main concepts, facts and methods from both chemical kinetics and statistical mechanics perspectives. Then, we develop a proper theoretical framework able to bridge statistical mechanics and chemical kinetics; the former can also serve as a proper tool for describing and investigating cybernetics, thus, as a syllogism, chemical kinetics and cybernetics become also related. The agreement of our framework with real data, carefully extrapolated from recent biological researches, covering the various standard behaviors in chemical kinetics, is also successfully checked. Finally, results and outlooks are discussed.

\section*{Results}

\subsection*{Collective behaviors in chemical kinetics}
Many polymers and proteins exhibit cooperativity, meaning that their ligands bind in a non-independent way: if, upon a ligand binding, the probability of further binding (by other ligands) is enhanced, like in the paradigmatic case of hemoglobin \cite{thompson}, the cooperativity is said to be positive, vice versa there is negative cooperativity when the binding of more ligands is inhibited \cite{kosh}, as for instance in some insulin receptors \cite{insulina1,insulina2} and most G-protein coupled receptors \cite{dativeri,rovira}.
Several mechanisms can be responsible for this effect: for example, if two neighbor docking sites on a polymer can bind charged ions, the electrostatic attraction/repulsion
may be the cause of a positive/negative cooperativity. However, the most common case is that the binding of a
ligand somehow modifies the structure of the hosting molecule, influencing the binding over the other
sites and this is the so-called allosteric mechanism \cite{rebek}.


Let us now formalize such behavior by considering a hosting molecule $P$ that can bind $N$ identical molecules $S$ on its structure; calling $P_j$ the complex of a molecule $P$ with $j \in [0,N]$ molecules attached, the reactions leading to the chemical equilibrium are the following
$$ S+ P_{j-1} \rightleftharpoons P_j,$$
hence the time evolution of the concentration of the unbounded protein $P_0$ is ruled by
\be
\frac{d[P_0]}{dt} = - K_{+1}^{(1)}[P_0][S]+K_{-1}^{(1)}[P_1],
\ee
where $ K_{+1}^{(1)},  K_{-1}^{(1)}$ are, respectively, the forward and backward rate constants for the state $j=1$, and their ratios define the {\em association constant} $K^{(1)} \equiv  K_{+1}^{(1)}  / K_{-1}^{(1)}$ and {\em dissociation constant} $\tilde{K}^{(1)} \equiv  K_{-1}^{(1)} / K_{+1}^{(1)}$.
Focusing on the steady state we get, iteratively,
$$
 K^{(j)}=\frac{[P_j]}{[P_{j-1}][S]}.
$$
Unfortunately, measuring $[P_j]$ is not an easy task and one usually introduces, as a convenient experimental observable, the average number $\bar{S}$ of substrates bound to the protein as
\be \label{eq:adair}
\bar{S} = \frac{[P_1]+2[P_2]+...+N[P_N]}{[P_0]+[P_1]+...+[P_N]} = \frac{K^{(1)}[S] + 2 \, K^{(2)}[S]^2+...+ N \, K^{(N)}[S]^N}{1+  K^{(1)}[S] +  K^{(1)}K^{(2)}[S]^2+...+ K^{(1)} \, K^{(N)}[S]^N},
\ee
which is the well-known Adair equation \cite{PCK}, whose normalized expression defines the {\em saturation function} $Y=\bar{S}/N$.
\newline
In a {\em non-cooperative} system, one expects independent and identical binding sites, whose steady states can be written as (explicitly only for $j=1$ and $j=2$ for simplicity)
\begin{eqnarray}
0 &=& -N K_{+}[P_0][S]-K_{-}[P_1], \ \ \ \ (j=1), \\
0 &=& -(N-1) K_+ [P_1][S] + 2K_- [P_2], \ \ \ \ (j=2),
\end{eqnarray}
where $K_+$ and $K_-$ are the rates for binding and unbinding on any arbitrary site.
\newline
Being $K \equiv K_+/K_-$ the intrinsic association constant, we get
\begin{eqnarray}
K &=& \frac{[P_1]}{N[P_0][S]}= \frac{K^{(1)}}{N},  \ \ \ \ (j=1), \\
K &=& \frac{2 [P_2]}{(N-1)[P_1][S]} = \frac{2K^{(2)}}{(N-1)}, \ \ \ \ (j=2),
\end{eqnarray}
and, in general, $K^{(j)}= (N-j+1)\, K / j$. Plugging this expression into the Adair equation (\ref{eq:adair}) we get
\be \label{MM}
\bar{S} = \frac{N K [S]}{1+ K [S]} \Rightarrow Y = \frac{K [S]}{1+K[S]},
\ee
which is the well-known Michaelis-Menten equation \cite{PCK}.

If interaction among binding sites is expected, the kinetics becomes far less trivial. Let us first sketch the limit case where intermediates steps can be neglected, that is
$$ [P_0]+ N[S] \rightleftharpoons [P_N],$$
then
\begin{eqnarray}
\bar{S} &=& \frac{N [P_N]}{[P_0]+[P_N]} = \frac{N K [S]^N}{1+ [S]^N}, \\
\bar{Y} &=& \frac{Y}{N} = \frac{K [S]^N}{1+ [S]^N}.
\end{eqnarray}
More generally, one can allow for  a degree of sequentiality  and write
\be
Y = \frac{K [S]^{n_H}}{1+[S]^{n_H}},
\ee
which is the well-known Hill equation \cite{PCK}, where $n_H$, referred to as Hill coefficient, represents the effective number of substrates which are interacting, such that for $n_H =1$ the system is said to be {\em non-cooperative} and the Michaelis-Menten law is recovered; for $n_H > 1$ it is {\em cooperative}; for $n_H \gg 1$ it is {\em ultra sensitive}; for $n_H < 1$ it is {\em anti cooperative}.
\newline
From a practical point of view, from experimental data for $Y([S])$, one measures $n_H$ as the slope of $\log(Y/(1-Y))$ versus $[S]$.
\newline

\subsection*{Mean-field statistical mechanics}

One of the best known statistical mechanics model is the mean-field Ising model, namely the Curie-Weiss model  \cite{ellis}. It describes the macroscopic behavior of a magnetic system microscopically  represented by $N$ binary spins, labeled by $i=1,2,...,N$, and whose state is denoted by $\sigma_i = \pm1$. In the presence of an external field $\mathbf{h}$ and being $\mathbf{J}$ the $N \times N$ symmetric  matrix encoding for pairwise interactions among spins, the (extensive, macroscopic)  internal energy associated to a the configuration $\{ \sigma \} = \{\sigma_1, \sigma_2,...,\sigma_N\}$ is defined as
\begin{equation}\label{eq:CW}
E(\{ \sigma \}|\mathbf{J},\mathbf{h}) = - \frac{1}{N} \sum_{i=1}^N \sum_{j<i}^{N} J_{ij} \sigma_i \sigma_j - \sum_{i=1}^N h_i \sigma_i.
\end{equation}
It is easy to see that the spin configurations leading to a lower energy are those where spins are aligned with the pertaining field, i.e. $\sigma_i h_i >0$, and pairs $(i,j)$ associated to positive (negative) coupling $J_{ij}$ are parallel (anti-parallel), i.e. $\sigma_i \sigma_j=1$ ($\sigma_i \sigma_j=-1$).
Notice that, in eq.~\ref{eq:CW}, we implicitly assumed that any arbitrary spin possibly interacts with any other.
This is a signature of the mean field approach which, basically, means that interactions among spins are long-range and/or that the time-scale of reactions is longer than the typical time for particles to diffuse, in such a way that each spin/particle actually sees any other. We stress that the mean field approximation also implies that the probability distribution $P(\{\sigma\})$ for the whole configuration is factorized into the product of the distribution for each single constituents, namely $P(\{\sigma\})=\prod_{i=1}^N P(\sigma_i)$, analogously to classical chemical kinetic prescriptions \cite{chay}.

For the analytic treatment of the system 
it is convenient to adopt a mesoscopic description where the phase space, made of all the $2^N$ possible distinct spin configurations, undergoes a coarse-graining and is divided into a collection $\mathcal{T}$ of sets, each representing a mesoscopic state of a given energy $E_k$, (we dropped the dependence on the parameters $\mathbf{J}, \mathbf{h}$ to lighten the notation). In this way, all the microscopic states belonging to the set $k \in \mathcal{T}$ share the same value of energy $E_k$, calculated according to (\ref{eq:CW}).
In order to describe the macroscopic behavior of the system through its microscopical degrees of freedom, we introduce a statistical ensemble $\rho \equiv \{ \rho_k \}_{k \in \mathcal{T}}$, meant as the probability distribution over the sets in $\mathcal{T}$; consequently, $\rho_i\geq 0$ and $\sum_k \rho_k =1$ must be fulfilled.
Accordingly, the internal energy and the entropy read as
\begin{eqnarray}
E (\rho)= \sum_{k \in \mathcal{T}} E_k \rho_k, \;\;\;\;
S (\rho)= - K_B \sum_ {k\in \mathcal{T}} \rho_k \log \rho_k,
\end{eqnarray}
where $K_B$ is the Boltzmann constant, hereafter set equal to $1$.
Being $\beta>0$ the absolute inverse temperature of the system, we define the free-energy
\be \label{eq:free}
F(\rho, \beta) = E (\rho) - \beta^{-1} S(\rho).
\ee
Notice that the minimum of $F$ ensures, contemporary, the minimum for $E$ and the maximum for $S$, hence it provides a definition for the thermodynamic equilibrium. As a consequence,  from eq.~\ref{eq:free} we calculate the derivative with respect to the probability distribution and require $\partial F / \partial \rho_i=0$; the solution, referred to as $\bar{\rho}$, reads as
\be \label{rho_eq}
\bar{\rho}_k = \frac{e^{-\beta E_k}}{Z(\beta)},
\ee
and it is called the Maxwell-Bolzmann distribution. The normalization condition implies $Z (\beta)= \sum_k e^{-\beta E_k}$ and this quantity is called ``partition function''.
We therefore have
\be
\bar{F}(E,\beta) = \inf_{\rho} F(E, \rho, \beta) \equiv F(E, \bar{\rho},\beta)= - \beta^{-1}\ln Z(\beta).
\ee
In general, given a function $f(\{\sigma\})$, its thermal average is $\langle f \rangle = \sum_{\{\sigma\}} f(\{\sigma\})e^{-\beta E(\{\sigma\})} / Z(\beta)$.

As this system is expected to display two different behaviors, an ordered one (at low temperature) and a disordered one (at high temperature), we introduce the magnetization $m = \sum_{i=1}^N \sigma_i /N$, which provides a primary description for the macroscopic behavior of the system. In particular, it works as the ``order parameter'' and it characterizes the onset of order at the phase transition between the two possible regimes. More precisely, as the parameters $\beta, \textbf{J}, \textbf{h}$ are tuned (here for simplicity $J_{ij} > 0, \forall i, j$), the system can be either disordered (i.e. paramagnetic), where spins are randomly oriented and $\langle m \rangle=0$, or ordered (i.e. ferromagnetic), where spins are consistently aligned and $\langle m \rangle \neq 0$. The phase transition, separating regions where one state prevails against the other, is a consequence of the collective microscopic interactions.
\newline
In a uniform system where $J_{ij} \equiv J, \forall i \neq j, J_{ii}=0$, and $h_i \equiv h, \forall i$, all spins display the same expected value, i.e. $\langle \sigma_i \rangle = \langle \sigma \rangle, \forall i$, which also corresponds to the average magnetization $\langle m \rangle$. Remarkably, in this case the free energy of the system can be expressed through $\langle m \rangle$ by a straightforward calculation \cite{amit} that yields
\be
 F(\beta,h)= - \beta^{-1} \ln 2 -\beta^{-1}  \ln \cosh \left[\beta(J \langle m \rangle +h)\right] +  \frac{1}{2} \langle m^2 \rangle,
\ee
whose extremization w.r.t. to $\langle m \rangle$ ensures again that thermodynamic principles hold and it reads off as
\be\label{eq:SC}
\langle m (\beta, J, h) \rangle =  \tanh[\beta (J \langle m \rangle + h)],
\ee
which is the celebrated Curie-Weiss self-consistency. By simply solving eq.~\ref{eq:SC} (e.g. graphically or numerically) the macroscopic behavior can be inferred.
Before proceeding, we fix $\beta =1$, without loss of generality as it can be reabsorbed trivially by $h \to \beta h = h$ and $J \to \beta J = J$.

In the non-interacting case ($J = 0$), eq.~\ref{eq:SC} gets $m(J=0, h) = \tanh (h)$, which reminds to an input-output relation for the system. When interactions are present ($J>0$), one can see that the solution of eq.~\ref{eq:SC} crucially depends on $J$. Of course, $ \langle m(J=0, h)\rangle < \langle m(J>0, h) \rangle$, due to cooperation among spins, and, more remarkably, there exists a critical value $J_c$ such that
 when $J \geq J_c$ the typical sigmoidal response encoded by $\langle m(J, h) \rangle$ possibly becomes a step function (a true discontinuity is realized only in the thermodynamic limit $N \to \infty$, while at finite $N$ the curve gets severely steep but still continuous).
Hence, to summarize, the Curie Weiss model exhibits two phases: A small-coupling phase where the system behaves paramagnetically and a strong-coupling phase where it behaves ferromagnetically. The ferromagnetic states are two, characterized by positive and negative magnetization, according to the sign of the external field.

\subsection*{Statistical mechanics and chemical kinetics}

In this section we develop the first part of our formal bridge and show how the Curie-Weiss model can be looked from a biochemical perspective. We will start from the simplest case of independent sites and later, when dealing with interacting sites, we will properly generalize this model in order to consistently include both positive and negative cooperativity.


\subsubsection*{The simplest framework: non interacting sites}
Let us consider an ensemble of elements (e.g. identical macromolecules, homo-allosteric enzymes, a catalyst surface), whose interacting sites are overall $N$ and labelled as $i=1,2,...,N$. Each site can bind one smaller molecule (e.g. of a substrate) and we call $\alpha$ the concentration of the free molecules ($[S]$ in standard chemical kinetics language). We associate to each site an Ising spin such that when the $i^{th}$ site is occupied $\sigma_i=+1$, while when it is empty $\sigma_i = -1$. A configuration of the elements is then specified by the set $\{ \sigma \}$.


First, we focus on non-collective systems, where no interaction between binding sites is present, while we model the interaction between the substrate and the binding site by an ``external field" $h$ meant as a proper measure for the concentration of free-ligand molecules, hence $h=h(\alpha)$. One can consider a microscopic interaction energy given by
\be\label{MM1}
E(\sigma,h)=-h\sum_{i=1}^N \sigma_i.
\ee
Note that writing $h\sum_i \sigma_i$ instead of $\sum_i h_i \sigma_i$ implicitly assumes homogeneity and thermalized reactions: each site displays the same coupling with the substrate and  they all had the time to interact with the substrate.

We can think at $h$ as the chemical potential
for the binding of substrate molecules on sites: When it is positive, molecules tend to bind to diminish energy, while when it is negative, bound molecules tend to leave occupied sites.
The chemical potential can be expressed as the logarithm of the concentration of
binding molecules and one can assume that the concentration is proportional to the ratio of the
probability of having a site occupied with respect to that of having it empty. In this simple
case, and in {\em all} mean-field approaches \cite{ellis}, the probability of each configuration is the product of the single independent probabilities of each site to be occupied and, applying the Maxwell-Boltzmann distribution $P(\sigma_i = \pm 1) = e^{\pm h}$, one finds
\be\label{ponte1}
\alpha \propto \frac{P(\sigma_i=+1)}{P(\sigma_i=-1)}= \frac{e^{+h}}{e^{-h}},
\ee
and we can pose
\be
h = \frac{1}{2}\log \alpha.
\ee
The mean occupation number (close to the magnetization in statistical physics) reads off as
$$
n(\{\sigma\})= \frac12 \sum_{i=1}^N(1+\sigma_i).
$$
Therefore, using ergodicity to shift $\bar{S}$ into $\langle S \rangle$ (see eq.~\ref{MM}), the saturation function can be written as
\be \label{eq:Y}
Y(\alpha)=\frac{\langle S(\{\sigma\})\rangle}{N}=\frac1N \sum_{\sigma} P(\{\sigma\})n(\{\sigma\})=\frac12(1+\langle m(h) \rangle).
\ee
By using eq.~\ref{eq:SC} we get
\be
Y(\alpha)=\frac12 [1+ \tanh (h)],
\ee
which, substituting $2 h = \log \alpha$, recovers the Michaelis-Menten behavior, consistently with the assumption of no interaction among binding sites ($J=0$) in eq. \ref{MM1}.

\subsubsection*{A refined framework: two-sites interactions}

We now focus on pairwise interactions and, seeking for a general scheme, we replace the fully-connected network of the original Curie-Weiss model by a complete bipartite graph: sites are divided in two groups, referred to as A and B, whose sizes are $N_A$ and $N_B$ ($N=N_A+N_B$), respectively. Each site in A (B) is linked to all sites in B (A), but no link within the same group is present. With this structure we mirror dimeric interactions \footnote{Note that for the sake of clearness, we introduced the simplest bipartite structure, which naturally maps dimeric interactions, but one can straightforwardly generalize to the case of an $n$-mer by an $n$-partite system and of course values of $\rho_A \neq \rho_B$ can be considered too. We did not perform these extensions because we wanted to recover the broader phenomenology with the smaller amount of parameters, namely $J,\alpha$ only.}, where a ligand belonging to one group interacts in a mean field way with ligands in the other group (cooperatively or competitively depending on the sign of the coupling, see below), and they both interact with the substrate. As a result, given the parameters $\mathbf{J}$ and $\mathbf{h}$, the energy associated to the configuration $\{\sigma\}$ turns out be
\be
\label{bizione}
E(\{ \sigma \}|\mathbf{J},\mathbf{h}) = -\frac{1}{(N_A+N_B)}\sum_{i=1}^{N_A}\sum_{j=1}^{N_B}J_{ij} \sigma_i \sigma_j - \sum_{i=1}^{N_A+N_B}h_i \sigma_i.
\ee
Some remarks are in order now.
First, we stress that in eq.~\ref{bizione} the sums run over all the binding sites. As we will deal with the thermodynamic limit ($N \to \infty$), this does not imply  that we model macromolecules of infinite length, which is somehow unrealistic. Rather, we consider $N$ as the total number of binding sites, localized even on different macromolecules, and the underlying mean-field assumption implies that binding sites belonging to the same group are all equivalent, despite some may correspond to the bulk and others to the boundaries of the pertaining molecule; such differences can be reabsorbed in an effective renormalization of the couplings. In this way the system, as a whole, can exhibit (anti) cooperative effects, as for instance shown experimentally in \cite{ackers92}.
\newline
Moreover, for the sake of clearness, in the following we will assume that couplings between sites belonging to different groups are all the same and equal to  $J$ and, similarly, $h_i=h$, for any $i$. This homogeneity assumption allows to focus on the simplest cooperative effects and can be straightforwardly relaxed.
\newline
We also notice that this two-groups model can mimic both cooperative and non-cooperative systems but, while for the former case bipartition is somehow redundant as qualitatively analogous results are obtained by adopting a fully-connected structure, for the latter case the underlying competitive interactions intrinsically require a bipartite structure.

Now, the two groups are assumed as equally populated, i.e., $N_A=N_B=N/2$, such that their relative densities are $\rho_A = N_A/N = \rho_B = N_B/N = 1/2$. The order parameter can be trivially extended as
\be
m_A = \frac{1}{N_A}\sum_{i=1}^{N_A} \sigma_i, \ \ \ m_B = \frac{1}{N_B}\sum_{j=1}^{N_B} \sigma_j,
\ee
and, according to statistical mechanics prescriptions, we minimize the free energy coupled to the cost function (\ref{bizione}) and we get, in the thermodynamic limit, the following self-consistencies
\begin{eqnarray}\label{selfpositive1}
\langle m_A \rangle &=& \tanh\left[ J \rho_B \langle m_B \rangle + h \rho_A \langle m_A \rangle  \right], \\ \label{selfpositive2}
\langle m_B \rangle &=& \tanh\left[ J \rho_A  \langle m_A  \rangle + h \rho_B \langle m_B \rangle \right].
\end{eqnarray}
Through eqs. \ref{selfpositive1} and \ref{selfpositive2}, the number of occupied sites can be computed as
\be
n_A(\{ \sigma \})=\sum_{i=1}^{N_A}\frac12 (1+\sigma_i) = N_A \frac{1+  m_A }{2}, \ \ n_B(\{ \sigma \})=\sum_{j=1}^{N_B}\frac12 (1+\sigma_j) = N_B \frac{1+ m_B }{2},
\ee
from which we get the overall binding isotherm
\begin{equation}\label{eq:Y2}
Y(\alpha)= \frac{\langle n_A (\alpha) \rangle + \langle n_B (\alpha) \rangle}{N}.
\end{equation}
From eqs.~\ref{selfpositive1}-\ref{eq:Y2} one can see that $Y(\alpha)$  fulfills the following free-energy minimum condition
\be\label{titra1}
Y(\alpha; J)=\frac12 \left \{ 1 + \tanh\left[J(2Y-1)+ \frac12 \log \alpha \right] \right \}.
\ee
This expression returns the average fraction of occupied sites corresponding to the equilibrium state for the
system. We are now going to study separately the two cases of positive ($J >0$) and negative ($J <0$) cooperativity.

\subsection*{The cooperative case: Chemical kinetics}

When $J>0$ interacting units tend to ``imitate'' each other. In this ferromagnetic context one can prove that the bipartite topology does not induce any qualitative effects: results are the same (under a proper rescaling) as for the Curie Weiss model; indeed, in this case one can think of bipartition as a particular dilution on the previous fully-connected scheme and we know that (pathological cases apart), dilution does not affect the physical scenario \cite{camboni,Agliari-EPL2011,Barra-PhysicaA2012,Sgrignoli-2010JStat}.
\begin{figure}[!ht]\label{fig:teor}
\centering
\includegraphics[width=0.55\textwidth]{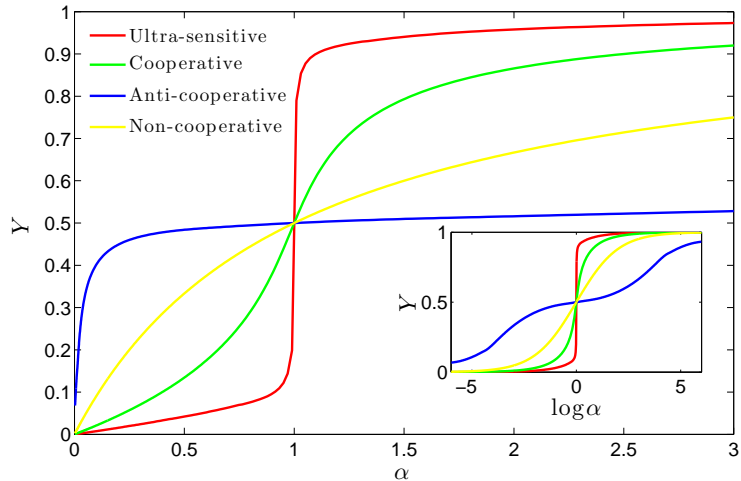}
\caption{Theoretical predictions of typical binding isotherms obtained from the statistical mechanics approach (see eq.~\ref{titra1}) are shown versus the substrate concentration $\alpha$ (main plot) and versus the logarithm of the concentration (inset). Different colors refer to different systems (hence different coupling strengths $J$), as explained by the legend. In particular, as $J$ is varied, all the expected behaviors emerge: ultra-sensitive for $J=2$, cooperative for $J=1/2$, anti-cooperative for $J=-1/2$, non-cooperative for $J=0$.}
\end{figure}

Differently from low-dimensional systems such as the linear Ising-chains, the Curie-Weiss model admits sharp (eventually discontinuous in the thermodynamic limit) transitions from an empty ($\langle m_A \rangle =\langle m_B \rangle =0$) to a completely filled ($\langle m_A \rangle =\langle m_B \rangle =1$) configuration as the field $h$ is tuned.
More precisely, eqs.~\ref{selfpositive1} and \ref{selfpositive2} describe a transition at $\alpha = 1$ and such a transition is second order ($Y$ changes continuously, but its derivative may diverge) when $J$ is smaller than the critical value $J_c = 1$, while it is first order ($Y$ has a discontinuity) when $J > J_c$.
The latter case is remarkable as a discontinuous behavior is experimentally well evidenced and at the basis of the so-called ultra-sensitive chemical switches \cite{germain}.
\newline
On the other hand, when $J \to 0$, the interaction term disappears  and we expect to recover Michaelis-Menten kinetics. In fact, eq.~\ref{titra1} can be rewritten as
\be\label{selfcoop}
Y(\alpha;J)= \frac{\alpha e^{2J (2Y-1)}}{1+\alpha e^{2J(2Y-1)}},
\ee
which, for $J=0$, recovers the Michaelis-Menten equation \footnote{Note that we do not lose generality when obtaining eq. (\ref{MM}) and not $Y(\alpha)=\alpha/(K+\alpha)$ (which is the more familiar MM expression) because we can rewrite the latter as $Y(\alpha)=K^{-1}\alpha/(1+K^{-1}\alpha)$ by shifting $\alpha \to \alpha K^{-1}$ and $h \to [\log (\alpha /K) ]/2$.} $Y(\alpha)= \alpha/(1+\alpha)$.
In this case there is no signature of phase transition as $Y(\alpha)$ is continuous in any of its derivatives.

In general, the Hill coefficient can be obtained as the slope of $Y(\alpha)$ in eq.~\ref{titra1} at the symmetric point $Y=1/2$, namely
\be\label{hill}
n_H =  \frac{1}{Y(1-Y)} \frac{\partial Y}{\partial \alpha}|_{Y=1/2}= \frac{1}{1-J},
\ee
where the role of $J$ is clear: a large $J$, i.e. $J$ close to $1$, implies a strong cooperativity and vice versa.
\newline
One step forward, as the whole theory is now described through the functions appearing in the self-consistency, we can expand them obtaining polynomials at all the desired orders, more typical of the standard route of chemical kinetics. In particular, expanding eq. \ref{selfcoop} at the first order in $J$ we obtain
\begin{equation}\label{eq:Ycoop}
Y(\alpha) \approx \frac{(1-J) \alpha + \alpha^2}{1+2(1-J)\alpha+ \alpha^2 },
\end{equation}
which is nothing but  the Adair equation (eq. $2$) as far as we set $J= (1-\sqrt{{K^{(1)}}^3 {K^{(2)}}}/2)$ and we rescale $\alpha \rightarrow \alpha/\sqrt{K^{(1)} K^{(2)}}$.
These results and, in particular, the expression in eq.~\ref{selfcoop} are shown in complete generality in figs.~$1$ and $2$.

The expression in eq.~\ref{selfcoop} can also be used to fit experimental data for saturation versus substrate concentration. Indeed, through an iterative fitting procedure, implied by the self-consistency nature of our theoretical expression, we can derive an estimate for the parameter $J$ and, from this, evaluate the Hill coefficient through eq.~\ref{hill}. As shown in fig.~$3$, fits are successful for several sets of experimental data, taken from different fields of biotechnology. The Hill coefficients derived in this way and the related estimates found in the literature are also in excellent agreement.

\begin{figure}[!ht]
\begin{center}
\includegraphics[width=12cm]{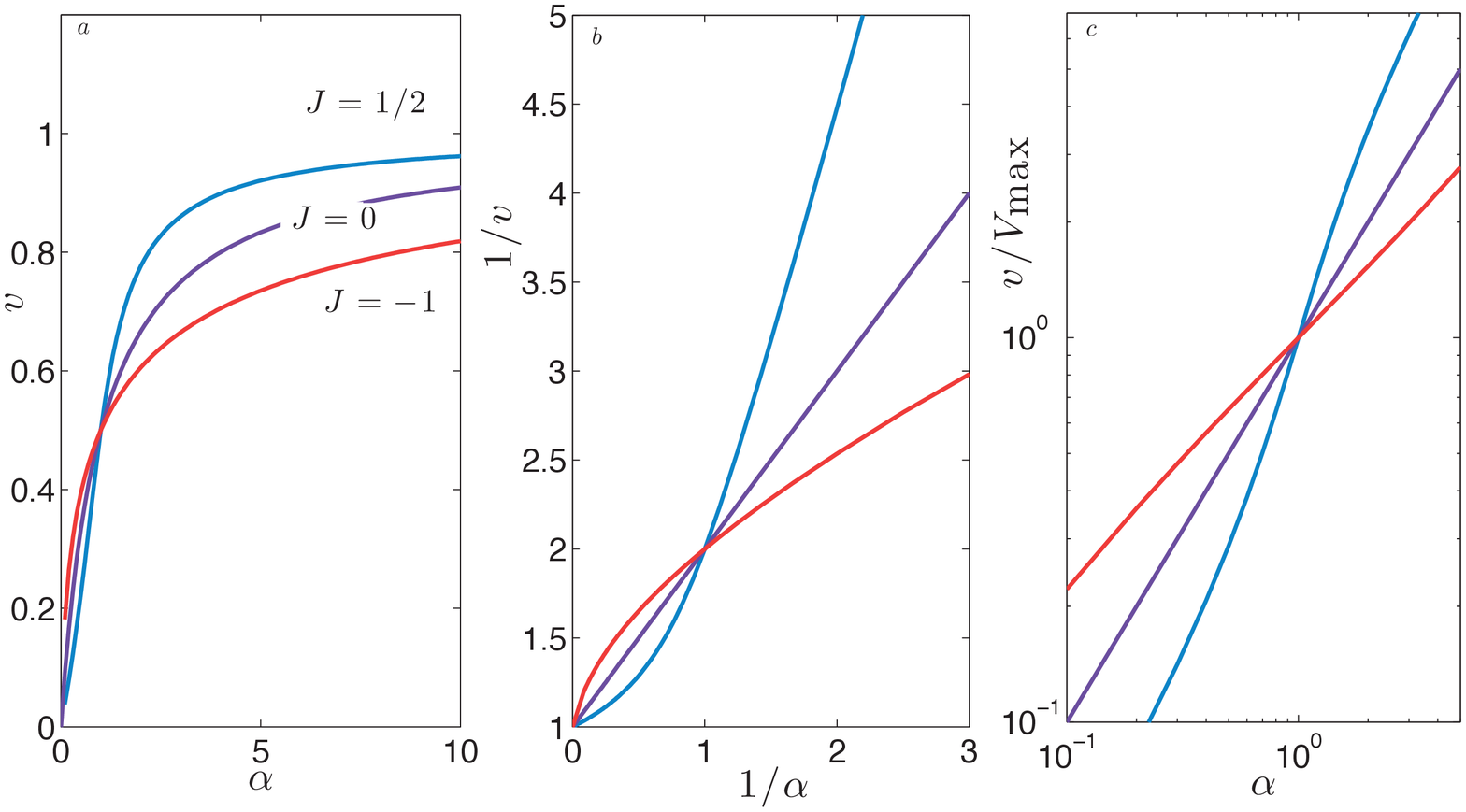}
\end{center}
\caption{Velocity of reactions versus substrate concentration $\alpha$. These plots have been included to show the full agreement between our theoretical outcomes and the results presented in  the celebrated paper by Levitzki and Koshland (see fig.$4$ in Ref. \cite{levitzki}). Different values of $n_H=1/2,1,2$, corresponding to $J=-1,0, 1/2$, are shown in different colors. Note that for this analysis there is complete proportionality between the reaction rates $v$ and the saturation curves $Y$ due to the law of mass action \cite{levitzki}.}
\label{fig:generalita}
\end{figure}

\begin{figure}[!ht]\label{fig:coop}
\begin{center}
\includegraphics[width=0.8\textwidth]{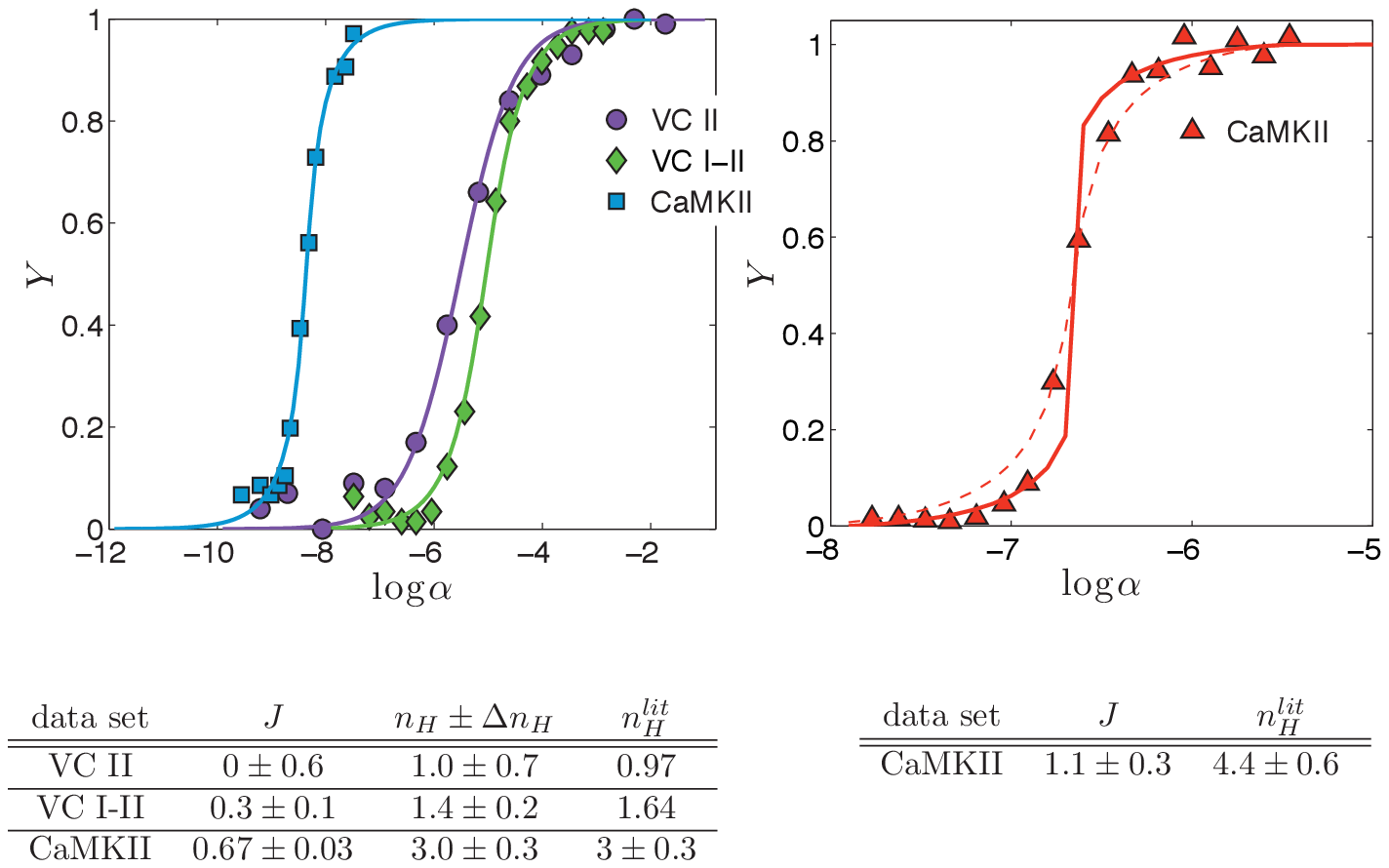}
\caption{These plots show comparison between data from recent experiments (symbols) and best-fits through statistical mechanics (lines).
Data refer to non-cooperative and positive-cooperative systems \cite{chao,mandal} (left panel) and an ultra-sensitive system \cite{bradshaw} (right panel). For the latter we report two fits: Dashed line is the result obtained by constraining the system to be cooperative but not-ultra-sensitive (that is, $J \leq 1$), while solid line is the best fit yielding $J \sim 1.1$, hence a ``first order phase transition" in the language of statistical mechanics. The relative goodness of the fits are $\chi^2_{coop}\sim 0.85$ and $\chi^2_{ultra} \sim 0.94$, confirming an ultra-sensitive behavior.
\newline
The tables in the bottom present the value of $J$ derived from the best fit and the resulting $n_H$; the estimate of the Hill coefficient taken from the literature is also shown.}
\end{center}
\end{figure}

\subsubsection*{Cooperative kinetics and cybernetics: Amplifiers and comparators}

Having formalized cooperativity through  statistical mechanics, we now want to perform a further translation in cybernetic terms. In particular, we focus on the electronic declination of cybernetics because this is probably the most practical and known branch. We separate the small coupling case ($J < J_c$, cooperative kinetics) from the strong coupling case ($J > J_c$, ultra-sensitive kinetics) and we mirror them to, respectively, the saturable operational amplifier and the analog-to-digital converter \cite{millman}. The plan is to compare the saturation curves (binding isotherms) in chemical kinetics with self-consistencies in statistical mechanics and transfer functions in electronics so to reach a unified description for these systems.

Before proceeding, we recall a few basic concepts. The core of electronics is the operational amplifier, namely a solid-state integrated circuit (transistor) which uses feed-back regulation to set its functions. In fig.~$4$ we show the easisest representation for operational amplifiers: there are two signal inputs (one positive received ($+$) and one negative received ($-$)), two voltage supplies ($V_{sat}, - V_{sat}$) and an output ($V_{out}$).
An {\em ideal} amplifier is the ``linear" approximation of the saturable one and essentially assumes that the voltage at the input collectors ($V_{sat}$ and $- V_{sat}$) is always at the same value so that no current flows inside the transistor, namely, retaining the obvious symbols of fig.~$4$, $i_+=i_-=0$ \cite{millman}. Obtaining its transfer function is straightforward as we can apply Kirchhoff law at the node $1$ to delete the afferent currents, hence $i_1 + i_2 + i_- = 0$. Then, assuming $R_1=1 \Omega$ (without loss of generality as only the ratio $R_2/R_1$ matters), in the previous equation we can pose $i_1 = - V_-$, $i_2 = (V_{out}-V_-)/R_2$ and  $i_-=0$ (because the amplifier is ideal). We can further note that $V_-=V_+$ and $V_+=V_{in}$ so to rewrite Kirchhoff law as $0=- V_{in}+(V_{out}-V_{in})/R_2$, by which the transfer function reads off as
\be \label{coop_V}
V_{out}= G \, V_{in}=(1+R_2)V_{in},
\ee
where $G=1+R_2$ is called ``gain'' of the amplifier.
\begin{figure}[!ht] \label{fig:negcoop}
\begin{center}
\includegraphics[width=8cm]{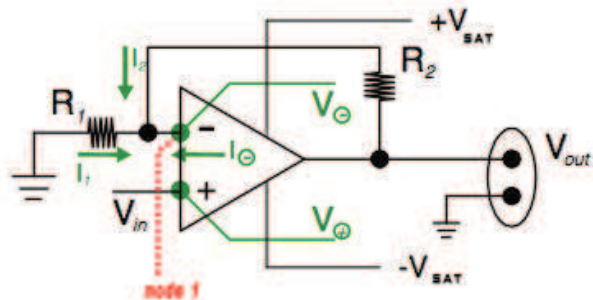}
\end{center}
\caption{Schematic representation of an operational amplifier.}
\label{fig:OA}
\end{figure}
Therefore, as far as $R_2 >0$, the gain is larger than one and the circuit is amplifying the input ($R_2 < 0$ is actually thermodynamically forbidden, suggesting that anti-cooperativity, from cybernetic perspective, must be accounted by the inverter configuration \cite{millman}, see next section).

Let us emphasize some structural analogies with ferromagnetic behaviors and cooperative kinetics. First, we notice that all these systems ``saturate''. Indeed, it is very intuitive to see that by applying a magnetic field $h>0$ to a collection of spins, they will (at least partially, depending on the noise level) align with the field, resulting in $\langle m(h) \rangle >0$. However, once reached the critical value $\tilde{h}$ such that $\langle m(\tilde{h}) \rangle =1$, any further increase in the strength of the field (i.e. any $h > \tilde{h}$) will produce no variations in the output of the system as all the spins are already aligned. Similarly, in reaction kinetics, once all the ligands of a given protein have bound to the substrate, any further growth in the substrate concentration will produce no net effect on the system. In the same way, given an arbitrary operational amplifier supplied with $V_{sat}$, then its output voltage will be a function of the input voltage $V_{in}$. However, there exists a critical input $\tilde{V}_{in}$ such that $V_{out}=V_{sat}$, and when input is larger than $\tilde{V}_{in}$ no further amplification is possible; the amplifier is then said to be ``saturated".
Of course, the sigmoidal shape of the hyperbolic tangent is not accounted by ideal amplifiers, yet for real amplifiers $\pm V_{sat}$ are upper bounds for the growth, hence recovering the expected behavior, as shown in the plot of fig.~$6 \, l$ \footnote{One may notice that the outlined amplifier is linear, while in chemical kinetics usually slopes are sigmoidal on lin-log plots. This is only a technical point, and to obtain the logarithmic amplifier it is enough to substitute $R_2$ with a diode to use the exponential scale of the latter \cite{millman}.}.
\newline
Moreover, we notice that the transfer function is an input/output relation, exactly as the equation for the order parameter $m$. In fact, the latter, for small values of the coupling $J$ (so to mirror ideal amplifier), can be written as (see eq.~$17$)
\be \label{coop_J}
\langle m \rangle \sim (1+J) h.
\ee
Thus, the external signal $V_{in}$ is replaced by the external field $h$, and  the voltage $V_{out}$ is replaced by the magnetization $\langle m \rangle$. By comparing eq.~\ref{coop_V} and eq.~\ref{coop_J} we see that $R_2$ plays as $J$, and, consistently, if $R_2=0$ the retroaction is lost (see fig.~$3$) and the gain is no longer possible. This is perfectly coherent with the statistical mechanics perspective, where, if $J=0$, spins do not mutually interact and no feed-back is allowed.

Analogously, in the chemical kinetics scenario, the Hill coefficient can be written as $n_H = 1/(1-J) \sim 1+J$, in the limit of small $J$ (namely for $J < J_c = 1$, which is indeed the case under investigation). Therefore, again, we see that if $J=0$ there is no amplification, and the kinetics returns the Michaelis-Menten scenario, while for positive $J$ we obtain amplification and a cooperative behavior. This leads to the conceptual equivalence
$$
\left( n_H = \frac{1}{1-J} \sim 1+J\right) \Leftrightarrow \left( G = 1+J \sim \frac{1}{1-J}\right),
$$
hence the Hill coefficient in chemical kinetics plays as the gain in electronics. This implicitly accounts for a quantitative comparison between amplification in electronics and in biological devices.

One step forward, if $J > J_c$ the equation $\langle m(h)\rangle$ becomes discontinuous in statistical mechanics just like the corresponding (ultra-sensitive) saturation curve in chemical kinetics: The analogy with cybernetics can still be pursued, but with analog-to-digital converters (ADCs), which are the corresponding limits of operational amplifiers.
\newline
The ADC, roughly speaking a switch, takes a continuous. i.e. analogue, input and has discrete (dichotomic in its basic implementation) states as outputs. The simplest ADC, namely flash converters, are built through cascades of voltage comparators \cite{millman}. A voltage comparator is sketched in fig.~$6 \, h$ and it simply ``compares" the incoming voltage values between the negative input and the positive one as follows:
Let us use as the negative input the ground ($V = 0$) as a reference value (to mirror one to one the equivalence with chemical kinetics or statistical mechanics we deal with only one input, namely the substrate concentration $\alpha$ in the former and the magnetic field $h$ in the latter). Then, if $V_{in}$ is positive the output will be $V_{sat} > 0$, vice versa, if $V_{in}$ is negative, the output will be $-V_{sat} < 0$ as reported in the plot in fig.~$6 \, m$, representing the ADC transfer function.
\newline
An ADC is simply an operational amplifier in an open loop (i.e. $R_2=\infty$), hence its theoretical gain is infinite. Coherently, this corresponds to values of $J \to 1$ that imply a theoretical divergence in the Hill coefficient, while, practically, reactions are referred to as ``ultra-sensitive'' already at $n_H \gg 1$. Consistently, as $J \to 1$ the curve $\langle m(h) \rangle$ starts to develop a discontinuity at $h=0$ (see Fig.$6$, panels $d$, $e$, $f$), marking the onset of a first order phase transition.
\newline
As a last remark, despite we are not analyzing these systems in the frequency domain in this first paper, we highlight  that, when using time-dependent fields, for instance oscillatory input signals,  full  structural consistency is preserved as all these systems display hysteresis effects at high enough frequencies of the input signal.

\subsection*{The anti-cooperative case: Chemical kinetics and cybernetics}
We can now extend the previous scheme for the description of a negative-cooperative system, by simply taking a negative coupling $J<0$.
Hence, eqs.~\ref{selfpositive1} and \ref{selfpositive2} still hold and we can analogously reconstruct $Y(\alpha)=(\langle m_A \rangle + \langle m_B \rangle) / N$ versus $\alpha$, whose theoretical outcomes are still shown in figs.~$1$ and $2$, and fit them against experimental results as shown in the plots of fig.~$5$.
\newline
Again, it is easy to check that there are two possible behaviors depending on the interaction strength $J$. If $J<J_c$, the two partial fractions $n_A, n_B$ are always equal, but when the interaction is larger than $J_c$, the two partial fractions are different in a region where the chemical potential $\log \alpha$ is around zero, as shown in fig.~$1$. In this region, due to the strong interaction and small chemical potential, it is more convenient for the system to fill sites on a subsystem and keep less molecules of ligands on the other subsystem.
The critical value of the chemical potential $|\log \alpha|$ depends on the interaction strength: it vanishes when the average interaction equals $ J_c $, and grows from this value on.
The region where the two fractions are different corresponds, in the magnetic models, to the anti-ferromagnetic phase.
\newline
When $J<J_c$, the binding isotherm, plotted as a function of the logarithm of concentration, has a form resembling the Michaelis-Menten curve, even if anti-cooperativity is at work.
Conversely, when $J>J_c$, in the region around $\alpha=1$ the curve has a concavity with an opposite sign with respect to the  Michaelis-Menten one.
In particular, there is a plateau around $\alpha=1$, which can be interpreted as the inhibition of the system, once it is half filled, towards further occupation.

Finally, in order to complete our analogy to electronics, let us consider the simplest bistable flip-flop \cite{millman}, built through two saturable operational amplifiers as sketched in fig.~$6 \, i$, such that the output of one of the two amplifiers is used as the inverted input of the other amplifiers, tuned by a resistor. This configuration, encoded in statistical mechanics by negative couplings among groups, makes the amplifiers reciprocally inhibiting because (and indeed they are called ''inverters'' in this configuration), for instance, a large output from the first amplifier (say A) induces a fall in the second amplifier (say B) and vice versa. Since each amplifier pushes  the other in the opposite state, there exist two stationary stable configurations (one amplifier with positive output and the other with negative output and vice versa). Thus, it is possible to assign a logical 0 (or 1) to one state and the other logical 1 (or 0) to the other state which can be regarded respectively as low concentration versus high concentration of bind ligands in chemical kinetics; negative or positive magnetization in ferromagnetic systems, low versus high output voltage of flip-flops in electronics. In this way, as the flip-flop can serve as an information storage device (in fact, the information (1 bit) is encoded by the output itself), the same feature holds also for the other systems.
The behavior of the two flip-flop transfer functions (one for each inverter) are also shown in
fig.~$6 \, n$, where the two (opposite) sigmoidal shapes are displayed versus the input voltage.
\newline
Still in fig.~$6$, those behaviors are compared with experimental data from biochemical anti-cooperativity and their statistical-mechanics best fits with an overall remarkable agreement.

\begin{figure}[!ht] \label{fig:negcoop}
\begin{center}
\includegraphics[width=12cm]{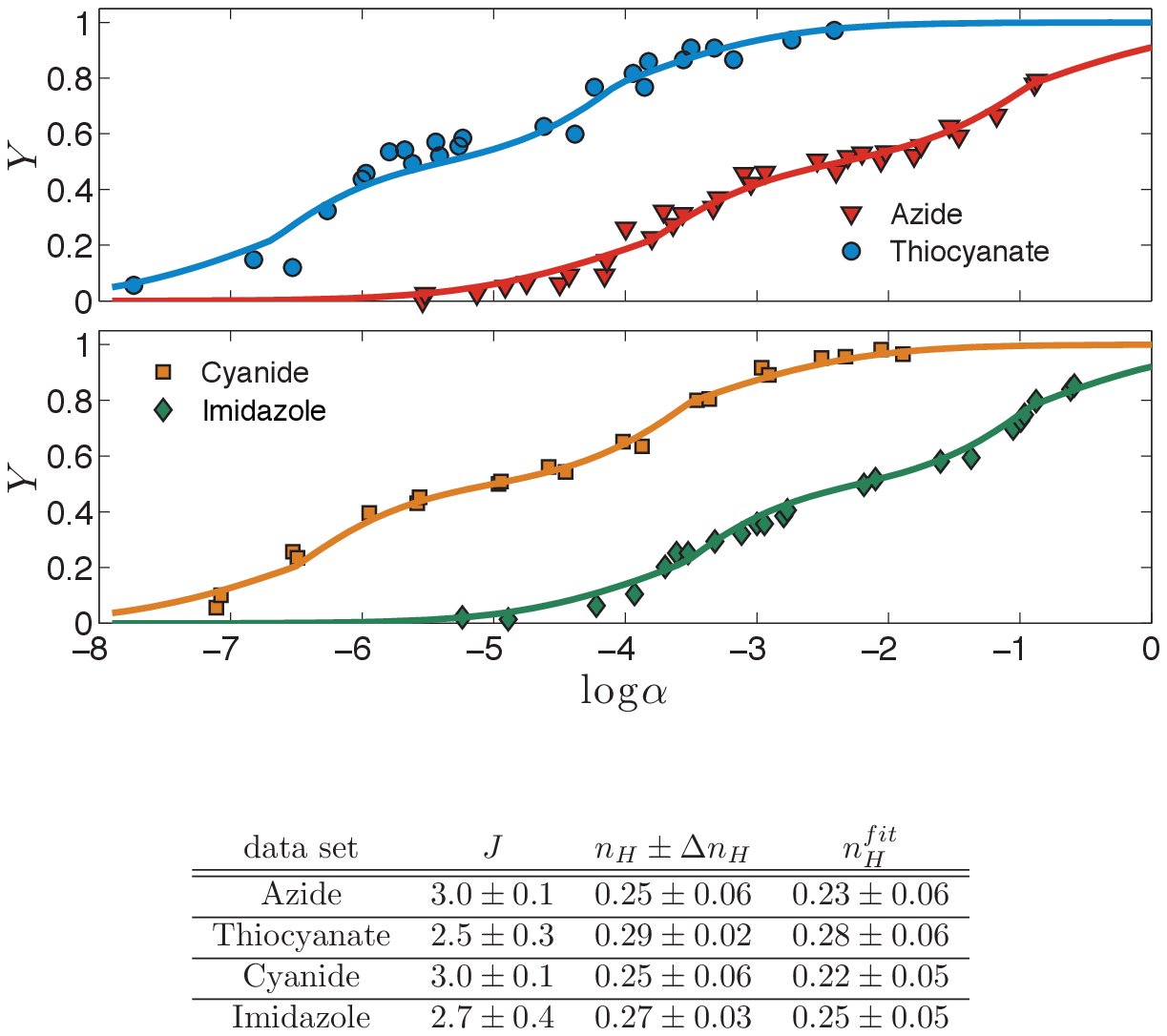}
\end{center}
\caption{Several sets of experimental data (symbols) \cite{bolognesi} are fitted by eq.~\ref{selfcoop} with minus sign (solid line). The values of $J$ corresponding to the best fits are shown in the table together with the related estimates for $n_H$ according to eq.~\ref{hill}. The estimates for $n_H$ obtained via standard Hill fit are also shown.}
\label{fig:BOLOGNESI}
\end{figure}

\begin{figure}[!ht]\label{fig:panels}
\begin{center}
\includegraphics[width=13cm]{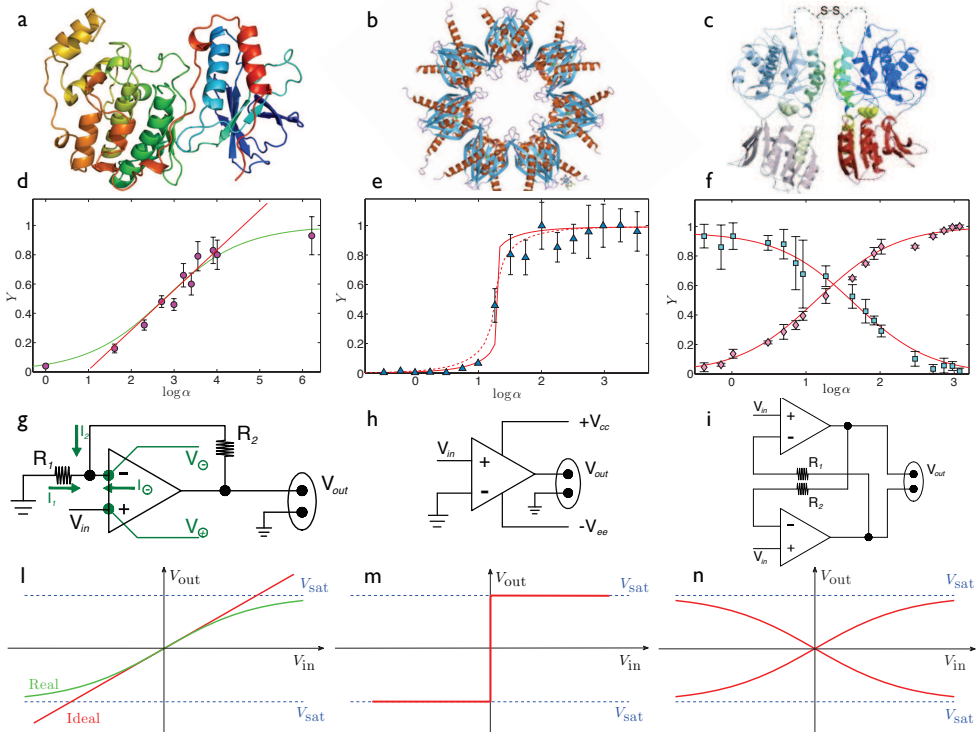}
\caption{This figure summarizes all the analogies described in the paper: In the first row, pictures of three biological systems exhibiting cooperativity, namely {\em Mitogen-activated protein kinase $14$} (positive cooperativity, panel $a$),  {\em $\textrm{Ca}^{2+}$ calmodulin dependent protein kinases II} (ultra-sensitive cooperativity, panel $b$), and {\em Synaptic Glutamate receptors} (negative cooperativity, panel $c$) are shown. The related saturation curves  (binding isotherms) are shown in the second row (panels $d$, $e$ and $f$, respectively), where symbols with the relative error-bars stand for real data taken from  \cite{solomatin,bradshaw,Suzuki-JBC2004} respectively and lines are best fits performed through the analytical expression in eq.~\ref{titra1}, obtained from statistical mechanics. The related best-fit parameters are $J=0.14$, $J=1.16$, $J=0.29$, respectively. Notice that in panel $d$ it is possible to see clearly the ``saturation" phenomenon as the first and the last experimental points are far from the linear fit (red line), while are perfectly accounted by the hyperbolic tangent predicted by statistical mechanics (green line), whose correspondence with saturation in electronics is represented in panel $l$. Notice further that in panel $e$, we compared the ultra-sensitive fit (solid line), with a simple cooperative fit (dashed line): at small substrate concentration the latter case does not match, within its variance, the data points (so accurately measured that error bars are not reported), while the former case is in perfect agreement with data points.
\newline
In the third row we sketch the cybernetic counterparts, i.e., the operational amplifier (panel $g$), represented as an inverted flip-flop mirroring the symmetry by which we presented the statistical mechanics framework (the standard amplifier is shown in fig.~$3$),  the analog-to-digital converter (panel $h$) and the flip-flop (panel $i$). The (theoretical) transfer functions corresponding to the circuits are finally shown in the fourth row (panels $l$, $m$ and $n$, respectively) for visual comparison with the second one.}
\end{center}
\end{figure}

\subsubsection*{Possible extensions: Heterogeneity and multiple binding sites.} \label{sec:extension}

Another point worth of being highlighted is the number of potential and straightforward extensions included in the statistical-mechanics modelization. In fact, as the literature of mean-field statistical-mechanics model is  huge, once self-consistencies are properly mapped into the saturation curves, one can perturb, generalize, or adjust the initial energy (cost) function and check the resulting effects.

As an example, we discuss chemical heterogeneity, which has been shown by recent experiments  \cite{solomatin,Macdonald-PNAS2008} to play a crucial role in equilibrium reaction rates. To include this feature in our theory we can replace $h \sum_i \sigma_i$ in eq. $18$ with $\sum_i h_i \sigma_i$ with $h_i$ drawn from, e.g., a Gaussian distribution
$$
P(h_i) \propto \exp\left[ -\frac{(h_i-h)^2}{2(1-a)^2} \right],
$$
such that for $a \to 1$ homogeneous chemical kinetics is recovered, while for $a \to 0$ we get standard Gaussian distribution $\mathcal{N}[h,1]$ for heterogeneity.
\newline
We fitted data from \cite{solomatin} through the self-consistencies obtained by either fixing $a=1$, or by taking $a$ as a free parameter; the results obtained are in strong agreement with the original one. In particular, the authors in \cite{solomatin} found a ratio $R$ between the ''real" Hill coefficient (assuming heterogeneity) and the standard (homogeneous) one as $R \sim 0.53$, while, theoretically we found $R \sim 0.57$ (and $a \sim 0.3$).
\newline
Further, we notice that $n_H$ grows with $|a-1|$, namely the higher the degree of inhomogeneity within the system and the smaller $n_H$, in agreement with several recent experimental findings \cite{Solomatin-Nature2011,Macdonald-PNAS2007}.

As a last example of possible extension, we discuss quickly also the multiple binding site case, which can be simply encoded, at least within the cooperative case, by considering an interaction energy of the $P$-spin type as
\be
E(\{\sigma\}|\bold{J},\bold{h}) =\frac{-1}{\sum_{i=1}^P N_i} \sum_{i_1}^{N_1}\sum_{i_2}^{N_2}...\sum_{i_P}^{N_P}J_{i_1,i_2,...,i_P}\sigma_{i_1}\sigma_{i_2}...\sigma_{i_P},
\ee
which results in multiple discontinuities for the binding isotherms as for instance happens when considering surfactants onto a polymer gel \cite{murase1999,chatterjee1996}, where the affinity of the surfactants to the gel is cooperatively altered by a conformational change of the polymer chains (and actually these systems show hysteresis with respect to the surfactant concentration, which is another typical feature of ``ferromagnetism").

\section*{Discussion}


In this work, we describe collective behaviors in chemical kinetics through mean-field statistical mechanics.  Stimulated by the successes of the latter in formalizing classical cybernetic subjects, as neural networks in artificial intelligence \cite{amit,hopfieldnature,MPV} or NP-completeness problems in  logic \cite{martin_monasson_zecchina_2001,neumann,sourlas_epl_1986},  we successfully tested the statistical mechanics framework as a common language to read from a cybernetic perspective chemical kinetic reactions, whose complex features are at the very basis of several biological devices.

In particular, we introduced an elementary class of models able to mimic possibly heterogeneous systems covering all the main chemical kinetics behaviors, namely ultra-sensitive, cooperative, anti-cooperative and non-cooperative reactions. Predictions yielded by such theoretical frame have been tested for comparison with experimental data taken from biological systems (e.g. nervous system, plasma, bacteria), finding overall excellent agreement. Furthermore, we showed that our analytical results recover all the standard chemical kinetics, e.g. Michaelis-Menten, Hills and Adair equations, as particular cases of this broader theory and confer to these a strong and simple physical background. Due to the presence of first order phase transition in statistical mechanics we offer a simple prescription to define a reaction as ultra-sensitive: Its best fit is achieved through a discontinuous function, whose extremization through other routes is not simple as e.g. least-squares can not be applied due to the discontinuity itself.

It is worth noticing that, despite we developed mean field techniques, hence we neglected any spatial structure, we get a direct mapping between statistical-mechanics and chemical kinetics formulas, in such a way that we can derive from the former a simple estimate for the Hill coefficient, namely for ''effective number" of interacting binding sites, in full agreement with experimental data and standard approaches.

One step forward, toward a unifying cybernetic perspective, we described a conceptual and practical mapping between kinetics of ultra-sensitive, cooperative and anti-cooperative reactions, with the behavior of analog-to-digital converters, saturable amplifiers and flip-flops respectively, highlighting how statistical mechanics can act as a common language between electronics and biochemistry. Remarkably, saturation curves in chemical kinetics mirror transfer functions of these three fundamental electronic devices which are the very bricks of robotics.

The bridge built here inspires and makes feasible several challenges and improvements in
biotechnology research. For instance, we can now decompose complex reactions into a sequence of elementary ones (modularity property \cite{winfree1}) and map the latter into an ensemble of interacting spin systems to investigate potentially hidden properties of the latter such as self-organization and computational capabilities (as already done adopting spin-glass models of neural networks \cite{prl}). Moreover, we can reach further insights in the development of better performing biological processing hardware, which are currently poorer than silico-made references. Indeed, from our equivalence between Hill and gain coefficients the more power of electronic devices is clear as $G$ can range over several orders of magnitude, while in chemical kinetics Hill coefficients higher than $n_H \sim 10$ are difficult to find. This, in turn, may contribute in developing a biological amplification theory whose fruition is at the very basis of biological computations \cite{winfree2,paun,berry}.
\newline
We believe that this is an important, intermediary, brick in the multidisciplinary research scaffold of biological complexity.

Lastly, we remark that this is only a first step: Analyzing, within this perspective, more structured biological networks as for instance the cytokine one at extracellular level or the metabolic one at intracellular level is still an open point and requires extending the mean-field statistical mechanics of glassy systems (i.e. frustrated combinations of ferro and antiferro magnets), on which we plan to report soon.

\section*{Acknowledges}

This works is supported by the FIRB grant $RBFR08EKEV$. Sapienza Universit\`{a} di Roma, INdAM through GNFM and INFN are acknowledged too for partial financial support.

\section*{Additional Information}

\subsection*{Author contribution statement}
The bridge between chemical kinetics and statistical mechanics has been built by all the authors.
\newline
The next bridge to cybernetics has been built by A.B.
\newline
E.A., A.D.B., G.U. performed data analysis and produced all the plots.
\newline
E.A., A.B., R.B. wrote the manuscript.

\subsection*{Competing financial interests}
The authors declare no competing financial interests.

\end{document}